\documentclass[conference]{IEEEtran}

\usepackage{cite}
\usepackage{amsmath,amssymb,amsfonts}
\usepackage{algorithmic}
\usepackage{graphicx}
\usepackage{textcomp}
\usepackage{xcolor}

\def\BibTeX{{\rm B\kern-.05em{\sc i\kern-.025em b}\kern-.08em
    T\kern-.1667em\lower.7ex\hbox{E}\kern-.125emX}}

\begin{document}

\title{DeepIntent: ImplicitIntent based Android IDS with E2E Deep Learning architecture}

\author{
\IEEEauthorblockN{Mohit Sewak}
\IEEEauthorblockA{\textit{Dept. of CS \& IS, Goa Campus} \\
\textit{BITS, Pilani, Goa, India}\\
p20150023@goa.bits-pilani.ac.in}
\and
\IEEEauthorblockN{Sanjay K. Sahay}
\IEEEauthorblockA{\textit{Dept. of CS \& IS, Goa Campus} \\
\textit{BITS, Pilani, Goa, India}\\
ssahay@goa.bits-pilani.ac.in}
\and
\IEEEauthorblockN{Hemant Rathore}
\IEEEauthorblockA{\textit{Dept. of CS \& IS, Goa Campus} \\
\textit{BITS, Pilani, Goa, India}\\
hemantr@goa.bits-pilani.ac.in}
}
\maketitle

\begin{abstract}
The \textit{Intent} in Android plays an important role in inter-process and intra-process communications. The \textit{implicit} Intent that an application could accept are declared in its manifest and are amongst the easiest feature to extract from an \textit{apk}. Implicit Intents could even be extracted online and in real-time. So far neither the feasibility of developing an Intrusion Detection System solely on implicit Intent has been explored, nor are any benchmarks available of a malware classifier that is based on implicit Intent alone. We demonstrate that despite Intent is implicit and well declared, it can provide very intuitive insights to distinguish malicious from non-malicious applications. We conducted exhaustive experiments with over 40 different end-to-end Deep Learning configurations of Auto-Encoders and Multi-Layer-Perceptron to create a benchmark for a malware classifier that works exclusively on implicit Intent. Using the results from the experiments we create an intrusion detection system using only the implicit Intents and end-to-end Deep Learning architecture. We obtained an area-under-curve statistic of \textbf{$0.81$}, and accuracy of $77.2\%$ along with false-positive-rate of \textbf{$0.11$} on Drebin dataset.
\end{abstract}


\section{Introduction} \label{sec:introduction}

Due to the open nature of its framework, a single Android smartphone can simultaneously run a large number of third party applications. This makes Android the most preferred
mobile platform globally amongst both its potential customers and
also malware developers and designers. Android uses
inter-process communication so that the components could be
used across processes, even those that belong to different applications. The inter-process communication mechanism
is also used as a gateway to access different sensitive services
in the Android framework. This communication mechanism
is driven by a late run-time binding object called 
\emph{Intent}.
\par
Intent is a messaging object that an app can use to request an action from another app component. When the calling and serving app components are part of the same app then it is called intra-process communication; and when these components reside in different apps, it is called inter-process communication. Intent can be used for requesting actions like \textit{Starting an activity}, \textit{Starting a service}, \textit{Delivering a broadcast}\cite{androidintent}, etc.
\par 
Android supports \textit{two} types of \textit{Intent}, namely \textit{Explicit} Intent and \textit{Implicit} Intent. In Explicit Intent, the calling app or app component knows the app or app component which can best provide the required activity/ service.  Hence the requesting app \textit{explicitly} specify which application will serve the required Intent. This is indicated by supplying just the target app's package name or a fully-qualified component class name. In \textit{Implicit} Intent, the calling app or app component may not be aware of the available application that could optimally service that desired Intent. Hence the requesting app just \textit{declare} a general action to perform, which allows a component from another app to service it. The Android system receives such Intent from the requesting app and finds target applications that could handle and service the requested Intent. The Android system uses the declarations made in the \textit{ AndroidManifest.xml} file of all installed applications to discover such potential target app. If there is only a single application available that could accept the requested implicit Intent, the Android system directly passes the associated message to that application; otherwise, it let the user choose from a list of all discovered potential applications that could serve the requested Implicit Intent.
\par
To empower this mechanism, an application that has the required functionality to serve a requested implicit Intent (and wish to receive requests from other applications to serve the same), needs to declare such Intents in the \textit{Intent Filter} of its manifest file. 

Therefore a declaration of an Intent in the intent filter within an app's manifest is useful only to inform the Android system that the specific app could accept and serve some particular implicit Intents from other requesting applications. Despite such declaration, a requested Intent would be routed to the declaring app only when such a request originates from any other app and the user manually choose this app as the preferred target from a list of competing candidate applications that could service the same Implicit Intent.
\par
Hence, such implicit and declared Intent may not provide as expressive features for detection of potential malicious nature. Richer insights could supposedly come from analysis of API calls, permissions, op-codes or even the activities that these apps conduct while serving the received implicit Intent request. Hence, not many IDS use such implicit Intents for malware detection. To the best of our information, no IDS has been explored or developed that use implicit Intents exclusively; and no-prior art has covered the effectiveness of Malware detection using exclusively such \textit{declared, implicit, Intents}. Implicit Intents could be extracted directly from an apk's manifest (along with permissions), and hence are amongst the easiest to extract features from an apk and could even be extracted in real-time without requiring any sandbox or 3rd party services. Hence such implicit Intent could be used for developing an easy to implement independent online IDS. But, to the best of our knowledge, no benchmark exists of any IDS/ malware-classifier that exclusively use Implicit Intent. 
\par
End-to-end (E2E) Deep Learning(DL) is a type of Deep Learning where the raw inputs are converted to the desired output/predictions using all DL models and obviates manual feature engineering. E2E DL has recently become very popular in almost all advanced applications of AI. 
This approach has been successfully applied to many actively researched problems ranging from image classification \cite{hinton-imagenet,socher2012convolutional} to neural-machine-translation \cite{joint-neural-translation} to search \cite{deep-search} and have delivered class-leading performance.
E2E DL architectures have shown good potential in the area of malware detection in Portable Executable (PE) \cite{malware-whole-exe, sewak-snpd, sewak-ares}. But in the case of Android malware classification DL has been used more conservatively and not in E2E manner. To the best of our information, this is the first work that uses an E2E DL approach for making an Android IDS.
\par
The major contributions of this paper are as follows:
\begin{itemize}
    \item Exploring the richness and intuitiveness of the insights that are provided by the Implicit Intents.
    \item Evaluating over $40$ E2E DL configurations of Auto Encoders (AE) and Multi-Layer Perceptron based Deep Neural Networks (MLP-DNN) in a structured manner and creating performance benchmarks for malware classification that use Implicit Intent as the only features.
    \item Designing an easy to implement online IDS named \textit{\textbf{DeepIntent}} that use and E2E DL implementation for both unsupervised feature engineering and supervised learning and use implicit Intents as the only features.
    \item Demonstrating that even with Implicit Intent alone, a very potent IDS can be developed and the resulting IDS could detect a malicious android application with an AUC of \textbf{$0.81$}, accuracy of $77.2\%$ along with a FPR of $0.11$.
\end{itemize}

The rest of the paper is organized as follows. Section \ref{sec:related-work}  discusses the related work on various static features and algorithms used for Android IDS. Section \ref{sec:data} explains the experimental dataset and section \ref{sec:intent-analysis} elaborates on insights from Implicit Intents extracted from the dataset. Section \ref{sec:experiments} describes over $40$ configurations of different deep learning architectures that were explored for making an effective and efficient IDS. In section \ref{sec:results} we discuss the experimental results and finally conclude the paper in section \ref{sec:conclusion}.

\section{Related Work} \label{sec:related-work}

Lately, the practice of developing an effective and easy to implement IDS for Android applications has evolved. Besides the conventional signature-based detection, a lot of machine learning-based detection techniques have also been attempted. The two essential aspects of any machine learning-based classifier are the input features and model algorithms. Static or dynamic/behavioural analysis are used to extract input features for the IDS. Since static features are extracted without executing the apps, these could be extracted at comparatively lower-cost and are thus efficient while ensuring broad coverage in real-time for an online IDS. 

\par
Amongst the static features used for an Android IDS, \textit{(Android) permission} are very popular. Au et al. developed PScout \cite{PScout} to analyze permissions from different versions of Android. They discovered that despite there is little redundancy in the specification of the $75$ permissions in Androids there is up to $22\%$ redundancy in the usage of the documented APIs. Lindorfer et al. developed an online service called ANDRUBIS \cite{ANDRUBIS}. They found that there is a difference in the number of permissions requested by benign versus malicious apps, where benign applications request far fewer permissions on average than their malicious counterparts. Though permissions are easy to extract, unlike op-codes, they do not offer semantically rich and useful insights. Therefore where some researchers like Puerta et al. \cite{Puerta-dalvic-opcode} and Li et al. \cite{li-permission} have focused exclusively on op-codes; others like Ashu et al. \cite{ashu-android}, and Idrees et al. \cite{Intent-Permission-Idrees} have used a combination of permissions and op-codes for making an effective malware detection system. Besides a combination of permissions and op-codes, some recent works have also explored a combination of other features with permission. For example, DroidSieve \cite{DroidSieve} combine permissions with other features like API calls, code structure, and the set of invoked components. 
\par
Recently the development of IDS using permissions with Intent has also gained traction. Wu et al. developed DroidMat, which uses permission, intent, and API Call for making a malware classifier \cite{DroidMat}. Similarly, Arp et al. used permission, intent, API call, and network address \cite{arp-drebin}. Ali et al. developed AndroDialysis that does not use permissions, but uses both implicit and explicit Intent for making an android malware detector \cite{AndroDialysis}. To the best of our knowledge, there exists no prior art which has explored making an Android IDS \textit{exclusively} using \textit{implicit Intent}.

In the area of model algorithms, various algorithms have been used for feature engineering and classification to develop a comprehensive Android IDS. Hemant et al. \cite{hemant-clustering} focused mainly on unsupervised learning and explored different types of clustering algorithms like K-means, Aggloromative BIRCH, DBSCAN and Gaussian Mixture Models for aiding a Random Forest-based IDS classifier; whereas Ashu et al. explored similar idea with functional tree \cite{ashu-android}. DroidMat \cite{DroidMat} used supervised models like k-nearest-neighbour and naive-bayes for building the malware classifier. Similarly, Arp et al. \cite{arp-drebin} utilized support vector machines (SVM) for the classification of malware. Rana et al. \cite{rana-tree} used multiple tree-based classifiers like a decision tree, random forest, gradient boosting, extremely randomized tree and random forest classifier on Android malware detection. Besides these, machine learning algorithms like Hidden Markov Model (HMM) \cite{pbmds}, Logistic Regression, and Bayesian Networks are also popular \cite{andromaly} for Android malware classification.

\section{Dataset and Pre-Processing} \label{sec:data}
In this section we discuss the malware dataset used, approach to collect and validate the complementing benign apps and the process to extract the Implicit Intents from these.

\subsection{Android Malware and benign applications}\label{subsec(data):malware}\label{subsec(data):benign}

We have used the \textit{`Drebin'} dataset\cite{arp-drebin} containing $5,560$ malware samples. The dataset contains samples from more than $20$ malware families.Drebin dataset is a superset of \textit{Android Malware Genome Project}\cite{zhou2012dissecting} which was proposed by Zhou et al. and contain more than $1,200$ malware samples. We downloaded more than $7,500$ Android applications from Google Play store. We used the online service of \textit{VirusTotal} for validating the downloaded application as benign. An Android app was labelled as benign if all the antivirus subscribed by \textit{VirusTotal} label it as non-malicious. We performed the exercise for all the downloaded applications and discarded the malicious ones. Finally, the benign dataset consists of $5,721$ benign Android apps.

\subsection{Extraction of Intents}\label{subsec(data):intent-extraction}

\textit{APKTOOL}, a $3^{rd}$ party reverse engineering tool, was used to disassemble the android applications (\textit{.apk} extension). It can decode an Android application into source files (including resources.arsc, classes.dex, XMLs etc.) and can also be used to recompile the application. Each application contains \textit{AndroidManifest(.xml)} file which describes essential information like App activities, services,  broadcast receivers etc. and are activated by intents. We extracted all the ($273$) Intent used by an application using a custom parser.

\section{Analysis of Implicit Intent}\label{sec:intent-analysis}
Since the malware and benign applications in the data set were mostly balanced, the Intent counts were directly analyzed without further normalization. Top $10$ Intents on the basis of total Intent count that are found in malicious applications (table \ref{table:Top20_Intent_Malicious}) and non-malicious samples (table \ref{table:Top20_Intent_Non-Malicious}) are tabulated. It was found that some Intents like \textit{BOOT\_COMPLETED}, \textit{ACTION\_POWER\_CONNECTED}, \textit{INPUT\_METHOD\_CHANGED}, \textit{UMS\_CONNECTED} etc. were comparatively more frequent in malicious apps than in non-malicious apps. For non-malicious apps, the most frequent \textit{unique} Intents were \textit{BROWSABLE}, \textit{SEARCH}, \textit{PICK}, \textit{INFO} etc. This difference between frequent Intents between the two classes of apps is a good indicator of the malicious modus-operandi of malicious applications.

\begin{table}[b]
\scriptsize
\begin{center}
\caption{Top-10 Intent used in malicious apps}
\begin{tabular}{|c|c|c|c|}
 \hline

\textbf{Type} & \textbf{Name} & \textbf{Count} & \textbf{Rank}\\
 \hline
action & MAIN & 7543 & 1\\
category & LAUNCHER & 6107 & 2\\
action & BOOT  COMPLETED & 4138 & 3\\
category & HOME & 1881 & 4\\
category & DEFAULT & 1524 & 5\\
action & USER  PRESENT & 607 & 6\\
action & VIEW & 496 & 7\\
action & ACTION  POWER  CONNECTED & 295 & 8\\
action & INPUT  METHOD  CHANGED & 293 & 9\\
category & SAMPLE  CODE & 249 & 10\\

\hline
\end{tabular}
\label{table:Top20_Intent_Malicious}
\end{center}
\end{table}

\begin{table}[htb]
\scriptsize
\begin{center}
\caption{Top-10 Intent used in non-malicious apps}
\begin{tabular}{|c|c|c|c|}
 \hline
\textbf{Type} & \textbf{Name} & \textbf{Count} & \textbf{Rank}\\
\hline
action & MAIN & 7388 & 1\\
category & DEFAULT & 6181 & 2\\
category & LAUNCHER & 6061 & 3\\
action & VIEW & 2281 & 4\\
category & BROWSABLE & 1486 & 5\\
action & BOOT  COMPLETED & 1097 & 6\\
action & PACKAGE  ADDED & 385 & 7\\
action & SEARCH & 325 & 8\\
action & SEND & 276 & 9\\
action & USER  PRESENT & 270 & 10\\

\hline
\end{tabular}
\label{table:Top20_Intent_Non-Malicious}
\end{center}
\end{table}

To explore this dissimilarity more objectively the difference in frequency of each Intent ($I_x$) was computed between the malicious($\mathcal{N}^{Mal}$) and non-malicious ($\mathcal{N}^{NonMal}$) applications and the vice-versa. The obtained difference was then normalized ($\mathbb{N(\mathcal{D})}_{I_x}$) by the average frequency of the specific Intent across both malicious and non-malicious apps as shown in equation \ref{eq:normalized-intent-difference}. Top 10 Intents by the average normalized difference between malicious over non-malicious ($\mathbb{N(\mathcal{D})}^{Mal}$) are shown in table \ref{table:Top20_Intent_Malicious_Over_Benign}. The difference between the frequency of some Intents like the \textit{DATA\_REMOVED}, \textit{GTALK\_CONNECTED}, \textit{INPUT\_METHOD\_CHANGED} etc was twice as much as the average frequency in the entire dataset. Intuitively understanding, this suggests that at least some malicious apps are more likely (than their non-malicious counterparts) to be dependant on the internet connection to accomplish their malicious actions. Also, some malicious apps may be engaged in malicious activities involving deletion of data of communication via installed communication apps.

\begin{equation}
    \mathbb{N(\mathcal{D})}^{Mal}_{I_x} = \frac{
    2*(
        \mathcal{N}^{Mal}_{I_x} - \mathcal{N}^{NonMal}_{I_x}
    )
    }{
    \mathcal{N}^{Mal}_{I_x} + \mathcal{N}^{NonMal}_{I_x}
    }
\label{eq:normalized-intent-difference}
\end{equation}

\begin{table}[b]
\scriptsize
\begin{center}
\caption{Top-10 Intent by difference between average presence in Malicious vs. Non-Malicious ones}

\begin{tabular}{|c|c|c|c|}
 \hline
\textbf{Type} & \textbf{Name} & \textbf{Count} & \textbf{Rank}\\
\hline
extra & DATA  REMOVED & 2 & 1\\
action & GTALK  CONNECTED & 2 & 1\\
action & INPUT  METHOD  CHANGED & 2 & 1\\
action & UMS  DISCONNECTED & 2 & 1\\
action & UMS  CONNECTED & 2 & 1\\
category & HOME & 1.61 & 6\\
action & BATTERY  OKAY & 1.61 & 6\\
category & SAMPLE  CODE & 1.57 & 8\\
action & BATTERY  LOW & 1.39 & 9\\
action & MEDIA  CHECKING & 1.33 & 10\\

\hline
\end{tabular}
\label{table:Top20_Intent_Malicious_Over_Benign}
\end{center}
\end{table}

\section{Experimental Setup \& Configurations} \label{sec:experiments}
Recently E2E DL architectures containing Stacked Auto Encoders (SAE) and MLP-DNN  have shown maturity \cite{sewak-overview-dl}, and these have demonstrated significant progress 
in the malware detection \cite{sewak-ares}. Therefore in this section, we experiment with similar E2E DL configurations, including SAE/AE and MLP-DNN.

\subsection{Finding the best AE configuration}\label{subsec:exp-ae}
SAE are type of AEs that are used to generate a lower dimension representation of original data while maintaining the non-linearity amongst the input features. 
The input dataset contains $273$ sparse features, many of them are NULL in many applications. Such input is not optimal to be used for supervised learning. By using SAE we reduce the number of features; convert them to dense representations from sparse; and preserve the expressive non-linearity in the relationship of the features.
Since the performance of MLP configurations are dependant upon the embeddings received from upstream AE, therefore we will first optimize the AE configuration. 

The validation-loss of the AE network and the AUC (Area Under Curve) of a downstream standardized MLP network were used as the performance criterion to appraise the different AE configurations.

In the first stage, we optimize No. of hidden layers, and No. of neurons in each of hidden layers and size of embedding layer for the SAE. Since these three hyper-parameters are inter-dependant, thus they cannot be evaluated in isolation. Hence several experiments with a different combination of these is multiple iterations were required while eliminating redundant configurations after each iteration. In the first iteration, we started evaluating different networks with a single hidden layer of varying hidden layer size and embedding dimension.

Table \ref{table:ae-single-layer-configuration} shows the AE configurations with single hidden layer of different sizes varying from $32$ neurons to $256$ neurons that were evaluated. As there were only $273$ unique Intents, therefore configurations involving layer size above $256$ neurons was not conducted. Similarly, as the embedding size was kept constant at $32$, therefore any hidden layer size less than $32$ neurons were not evaluated. The hidden layer size that demonstrated comparable performance with the least number of neurons was fixed as the upper cap for evaluating any configuration with more than one hidden layer in other iterations.

\begin{table}[b]
\scriptsize
\begin{center}
\caption{AE: Single Hidden Layer Configurations}
\scriptsize
\begin{tabular}{|c|c|c|c|c|}
\hline
\textbf{Conf. ID} & \textbf{\begin{tabular}[c]{@{}c@{}}Hidden \\ Layer Size\end{tabular}} & \textbf{\begin{tabular}[c]{@{}c@{}}Embedding\\ Dimension\end{tabular}} & \textbf{AE Val. Loss} & \textbf{MLP AUC} \\ \hline
1                 & [256]                                                                      & 256                                                                    & 0.016                 & 0.79             \\ 
2                 & [256]                                                                      & 128                                                                    & 0.017                 & 0.50             \\ 
3                 & [256]                                                                      & 64                                                                     & 0.017                 & 0.79             \\ 
4                 & [256]                                                                      & 32                                                                     & 0.018                 & 0.79             \\ 
5                 & [128]                                                                      & 128                                                                    & 0.017                 & 0.79             \\ 
6                 & [128]                                                                      & 64                                                                     & 0.018                 & 0.79             \\ 
7                 & [128]                                                                      & 32                                                                     & 0.018                 & 0.79             \\ 
8                 & [64]                                                                       & 64                                                                     & 0.018                 & 0.79             \\ 
9                 & [64]                                                                       & 32                                                                     & 0.018                 & 0.79             \\ 
10                & [32]                                                                       & 32                                                                     & 0.018                 & 0.50             \\ 
11                & [32]                                                                       & 16                                                                     & 0.019                 & 0.78             \\ \hline
\end{tabular}
\label{table:ae-single-layer-configuration}
\end{center}
\end{table}

Interestingly, during the first stage, some configurations demonstrated a sudden drop in performance; and configurations that had just a higher and just lower No. of neurons both had better performance. To ensure that this phenomenon was not just because of some trivial random No. initialization, these configurations were re-evaluated with a different random No. seed; but their performance did not change much, and hence these configurations along with some other were dropped in the next iteration.

In the second iteration, AE configurations with up to two hidden layers were evaluated. The sub-optimal performance of a DL network on a training dataset could be because of data complexity or non-linearity or both. Since the requirement of more than one hidden layer over the size of a hidden layer is more indicative of non-linearity, so layer sizes above $128$ neurons were not evaluated. As shown in table \ref{table:ae-two-layer-configuration}, in this iteration we evaluated two-hidden-layer configurations where the size of the first hidden layer could range between $128 \geq AE_{H1} \geq 32$ and the size of the second hidden layer could range between $AE_{H1} \geq AE_{H2} \geq 32$. Since in SAE the purpose is to compress the data dimension cardinality gradually, hence above constraint is required.

\textit{\begin{table}[htb]
\scriptsize
\begin{center}
\caption{AE: Two-Hidden-Layer Configurations}
\begin{tabular}{|c|c|c|c|c|}
\hline
\textbf{Conf. ID} & \textbf{\begin{tabular}[c]{@{}c@{}}Hidden \\ Layer Size\end{tabular}} & \textbf{\begin{tabular}[c]{@{}c@{}}Embedding \\ Dimension\end{tabular}} & \textbf{AE Val. Loss} & \textbf{MLP AUC} \\ \hline
14                & {[}128, 128{]}                                                        & 32                                                                      & 0.017                 & 0.5              \\
15                & {[}128, 64{]}                                                         & 32                                                                      & 0.017                 & 0.79             \\
16                & {[}128, 32{]}                                                         & 32                                                                      & 0.017                 & 0.5              \\
17                & {[}64, 64{]}                                                          & 32                                                                      & 0.017                 & 0.5              \\
18                & {[}64, 32{]}                                                          & 32                                                                      & 0.017                 & 0.5              \\
19                & {[}32, 32{]}                                                          & 32                                                                      & 0.017                 & 0.5              \\ \hline
\end{tabular}
\label{table:ae-two-layer-configuration}
\end{center}
\end{table}}

In the $3^{rd}$ iteration of AE configuration optimization, we evaluated multiple configurations with three hidden layers of varying sizes. The upper limit for the No. of layers in the first hidden layer the constraints for the same in any subsequent hidden layers were kept the same as in the $2^{nd}$ iteration. Table \ref{table:ae-3-layer-configuration} shows the various configurations of AE evaluated in the this iteration. Since none of these configurations performed as well as some other configurations with only two hidden layers as evaluated in earlier iteration, therefore configurations with more than three hidden layers were not evaluated.

\begin{table}[b]
\scriptsize
\begin{center}
\caption{AE: Three-Hidden-Layer Configurations}
\begin{tabular}{|c|c|c|c|c|}
\hline
\textbf{Conf. ID} & \textbf{\begin{tabular}[c]{@{}c@{}}Hidden \\ Layer Size\end{tabular}} & \textbf{\begin{tabular}[c]{@{}c@{}}Embedding \\ Dimension\end{tabular}} & \textbf{AE Val. Loss} & \textbf{MLP AUC} \\ \hline
20                & {[}128, 64, 64{]}                                                     & 32                                                                      & 0.017                 & 0.5              \\
21                & 128, 64, 32{]}                                                        & 32                                                                      & 0.017                 & 0.77             \\
22                & {[}128, 32, 32{]}                                                     & 32                                                                      & 0.017                 & 0.5              \\ \hline
\end{tabular}
\label{table:ae-3-layer-configuration}
\end{center}
\end{table}

\subsection{Finding the best MLP DNN configuration}\label{subsec:exp-mlp}
From the analysis of the experiments conducted in section \ref{subsec:exp-ae}, we determine the optimal AE configuration. Using the discovered optimal AE configuration we conducted multiple experiments to select an optimal configuration of the MLP DNN as well. The criterion for evaluating MLP DNN configuration we use the AUC statistic, along with the accuracy scored on the validation data. The validation accuracy is computed when the class as derived from the class probabilities using a threshold of $0.5$ for distinguishing between the two classes.

In the next stage, we conduct experiments with different numbers of hidden layers and varying sizes of each of these hidden layers for the MLP-DNN network. We have used similar constraints and shortlisting criteria as we did in the case of AE. We evaluated different MLP DNN configurations with $1$ to $6$ No. of hidden layers and with $32$ to $256$ No. of neurons in the hidden layers. Using the shortlisting criterion as indicated earlier, we were able to obviate evaluation of many configurations that were less likely to produce better results as compared to similar configurations that also has fewer number of hidden layers. The reduced number of experiments that were mandated as per the shortlisting subsequent experimentation criterion are as indicated in table \ref{table:mlp-hidden-layer-configuration}.

\begin{table*}[htb]
\scriptsize
\begin{center}
\caption{MLP: Experiments for finding the best number and size of hidden layers}
\begin{tabular}{|c|c|c|c|c|c|c|c|c|}
\hline
\textbf{Conf. ID} & \textbf{\begin{tabular}[c]{@{}c@{}}MLP Hidden\\ Layer Size\end{tabular}} & \textbf{MLP AUC} & \textbf{\begin{tabular}[c]{@{}c@{}}Accuracy\\ (Th=0.5)\end{tabular}} & \textbf{\begin{tabular}[c]{@{}c@{}}FPR\\ (Th=0.5)\end{tabular}} & \textbf{\begin{tabular}[c]{@{}c@{}}Best\\ Accuracy\end{tabular}} & \textbf{\begin{tabular}[c]{@{}c@{}}FPR @\\ Best Accuracy\end{tabular}} & \textbf{\begin{tabular}[c]{@{}c@{}}Accuracy @ \\Best F1\end{tabular}} & \textbf{\begin{tabular}[c]{@{}c@{}}FPR @ \\ Best F1\end{tabular}} \\ \hline

23 & [256] & 0.773 & 0.738 & 0.131 & 0.741 & 0.116 & 0.735 & 0.181\\
24 & [128] & 0.772 & 0.737 & 0.133 & 0.739 & 0.115 & 0.733 & 0.185\\
25 & [64] & 0.775 & 0.737 & 0.133 & 0.74 & 0.117 & 0.737 & 0.172\\
26 & [32] & 0.763 & 0.736 & 0.151 & 0.74 & 0.152 & 0.74 & 0.152\\
27 & [32, 32] & 0.778 & 0.736 & 0.137 & 0.74 & 0.164 & 0.734 & 0.19\\
28 & [64, 64] & 0.777 & 0.74 & 0.152 & 0.745 & 0.138 & 0.744 & 0.158\\
29 & [128, 128] & 0.781 & 0.744 & 0.141 & 0.746 & 0.151 & 0.746 & 0.151\\
30 & [256, 256] & 0.775 & 0.743 & 0.151 & 0.744 & 0.15 & 0.744 & 0.153\\
31 & [128, 128, 128] & 0.788 & 0.748 & 0.112 & 0.752 & 0.133 & 0.751 & 0.144\\
32 & [64, 64, 64] & 0.788 & 0.744 & 0.14 & 0.752 & 0.111 & 0.746 & 0.162\\
33 & [128, 128, 128, 128] & 0.501 & 0.514 & 0 & 0.514 & 0 & 0.486 & 0.999\\
34 & [64, 64, 64, 64] & 0.792 & 0.752 & 0.111 & 0.758 & 0.117 & 0.757 & 0.129\\
35 & [128, 128, 64, 64] & 0.788 & 0.743 & 0.116 & 0.751 & 0.136 & 0.749 & 0.15\\
36 & [64, 64, 64, 64, 64] & 0.788 & 0.74 & 0.126 & 0.75 & 0.137 & 0.745 & 0.161\\
37 & [64, 64, 64, 64, 64, 64] & 0.5 & 0.514 & 0 & 0.514 & 0 & 0.486 & 1\\
\hline
\end{tabular}
\label{table:mlp-hidden-layer-configuration}
\end{center}
\end{table*}

Next explored the effect of increasing the number of epochs for training. As shown in table \ref{table:mlp-increased-epochs}, we increased the number of epochs from $100$ (default in all previous experiments) to $1000$. Performance of configurations with more than $1000$ epochs were not evaluated as the training accuracy and loss plots obtained from the above configurations demonstrated convergence. 

\begin{table}[htbp]
\scriptsize
\begin{center}
\caption{MLP: Effect of increasing the number of Epochs}
\begin{tabular}{|c|c|c|c|c|c|}
\hline
\textbf{\begin{tabular}[c]{@{}c@{}}Conf.\\ ID\end{tabular}} & \textbf{\begin{tabular}[c]{@{}c@{}}MLP Hidden \\ Layer Size\end{tabular}} & \textbf{\begin{tabular}[c]{@{}c@{}}MLP\\ Epochs\end{tabular}} & \textbf{\begin{tabular}[c]{@{}c@{}}MLP\\ AUC\end{tabular}} & \textbf{\begin{tabular}[c]{@{}c@{}}Accuracy\\ (Th=0.5)\end{tabular}} & \textbf{\begin{tabular}[c]{@{}c@{}}FPR\\ (Th=0.5)\end{tabular}} \\ \hline
34                                                          & {[}64, 64, 64, 64{]}                                                      & 100                                                           & 0.792                                                      & 0.737                                                                & 0.111                                                           \\
38                                                          & {[}64, 64, 64, 64{]}                                                      & 1000                                                          & 0.802                                                      & 0.738                                                                & 0.101                                                           \\ \hline
\end{tabular}
\label{table:mlp-increased-epochs}
\end{center}
\end{table}

The optimizer algorithm plays a significant role in determining both the effectiveness (as determined by performance indicators) and the efficiency of training (as indicated from the computational time). 
We experimented with some of the popular optimizer algorithms in DL \cite{sewak-overview-dl} next. The default optimizer algorithm used so far was \textit{adadelta}. Besides, we also experimented with the \textit{adam} and \textit{rmsprop} optimizers keeping other parameters same as indicated in table \ref{table:mlp-optimizers}.

\begin{table}[htb]
\scriptsize
\begin{center}
\caption{MLP: Effect of change of Optimizer algorithm}
\begin{tabular}{|c|c|c|c|c|c|c|}
\hline
\textbf{\begin{tabular}[c]{@{}c@{}}Conf.\\ ID\end{tabular}} & \textbf{\begin{tabular}[c]{@{}c@{}}MLP Hidden\\ Layer Size\end{tabular}} & \textbf{\begin{tabular}[c]{@{}c@{}}MLP\\ Ep-\\ ochs\end{tabular}} & \textbf{\begin{tabular}[c]{@{}c@{}}MLP\\ Opti-\\ mizer\end{tabular}} & \textbf{\begin{tabular}[c]{@{}c@{}}MLP\\ AUC\end{tabular}} & \textbf{\begin{tabular}[c]{@{}c@{}}Acc\\ (Th\\ =0.5)\end{tabular}} & \textbf{\begin{tabular}[c]{@{}c@{}}FPR\\ (Th\\ =0.5)\end{tabular}} \\ \hline
38                                                          & {[}64, 64, 64, 64{]}                                                     & 1000                                                              & adadelta                                                             & 0.802                                                      & 0.768                                                              & 0.101                                                              \\
39                                                          & {[}64, 64, 64, 64{]}                                                     & 1000                                                              & adam                                                                 & 0.806                                                      & 0.764                                                              & 0.118                                                              \\
40                                                          & {[}64, 64, 64, 64{]}                                                     & 1000                                                              & rmsprop                                                              & 0.814                                                      & 0.772                                                              & 0.11                                                               \\ \hline
\end{tabular}
\label{table:mlp-optimizers}
\end{center}
\end{table}

Another hyper-parameter closely associated with both effectiveness and efficiency is the batch-size. The default batch size used until stage $6$ was $1024$. Hence for $\sim 7700$ training records ($~11000\times0.7$ training split), we made around 8 \textit{steps} of training passes in each epoch. 
In the next iteration we experimented with different batch-sizes as shown in table \ref{table:mlp-batch-size}. The choice of the optimizer algorithm, along with the optimizer`s associated hyper-parameters, like learning-rate ($\eta$) has a role to play in determining the optimal batch-size. Hence we had first finalized on the optimizer choices earlier before evaluating the different batch-sizes in this iteration.

\begin{table}[ht]
\scriptsize
\begin{center}
\caption{MLP: Effect of change of Batch Size}
\begin{tabular}{|c|c|c|c|c|c|c|}
\hline
\textbf{\begin{tabular}[c]{@{}c@{}}Conf.\\ ID\end{tabular}} & \textbf{\begin{tabular}[c]{@{}c@{}}MLP\\ Hidden\\ Layer Size\end{tabular}} & 
\textbf{\begin{tabular}[c]{@{}c@{}}MLP\\ Batch\\ Size\end{tabular}} & \textbf{\begin{tabular}[c]{@{}c@{}}MLP\\ Opti-\\ mizer\end{tabular}} & \textbf{\begin{tabular}[c]{@{}c@{}}MLP\\ AUC\end{tabular}} & \textbf{\begin{tabular}[c]{@{}c@{}}Acc\\ (Th\\ =0.5)\end{tabular}} & \textbf{\begin{tabular}[c]{@{}c@{}}FPR\\ (Th\\ =0.5)\end{tabular}} \\ \hline
40                                                          & \begin{tabular}[c]{@{}c@{}}{[}64,64,64,64{]}\end{tabular}                                                                                & 1024                                                                & rmsprop                                                              & 0.814                                                      & 0.772                                                              & 0.11                                                               \\
41                                                          & \begin{tabular}[c]{@{}c@{}}{[}64,64,64,64{]}\end{tabular}                                                                                & 2048                                                                & rmsprop                                                              & 0.803                                                      & 0.764                                                              & 0.107                                                              \\
42                                                          & \begin{tabular}[c]{@{}c@{}}{[}64,64,64,64{]}\end{tabular}                                                                                & 512                                                                 & rmsprop                                                              & 0.794                                                      & 0.755                                                              & 0.136                                                              \\ \hline
\end{tabular}
\label{table:mlp-batch-size}
\end{center}
\end{table}

\section{Results \& Discussion} \label{sec:results}

In section \ref{subsec:exp-mlp} we cover the experiment with multiple configurations of MLP DNN. The criteria for evaluating MLP DNN are the AUC statistic and the accuracy obtained on validation data (at prediction probability cut-off of 0.5). As shown in table \ref{table:mlp-hidden-layer-configuration} and \ref{table:best-configuration} we also compute the validation accuracy and FPR using different cut-off thresholds of prediction probabilities, viz.
\begin{itemize}
    \item the threshold cut-off of 0.5 (default for comparison),
    \item the threshold that gives the best accuracy (and associated FPR) even at a compromise of FPR,
    \item and third an accuracy and associated FPR that provides the best balance between the precision and recall as determined by the `F1 measure'. 
\end{itemize}

\begin{figure}[t]
     \vspace{-2mm}
    \caption{ROC Plot: MLP Conf. ID 40}
    \centering
    \includegraphics[width=3.4in, height=1.8in]{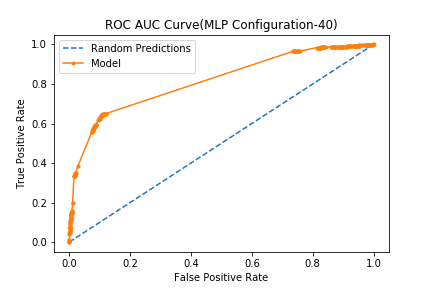}
    \label{fig:mlp-roc}
     \vspace{-2mm}
\end{figure}

We found that the MLP DNN with $4$ hidden layers, each with $64$ units, when trained with a batch-size of $1024$ records over $1000$ epochs using the `\textit{rmsprop}' optimizer, gave the best results. It produced an \textbf{AUC statistic of $0.814$} with an \textbf{accuracy of $77.2\%$} and an \textbf{FPR of $0.11$} on the validation data. This configuration is indicated by Conf. ID 40, and its details are shown in table \ref{table:best-configuration}. 
The ROC plot obtained at the different threshold for prediction probability cut-offs for this configuration are shown in figure \ref{fig:mlp-roc}.

An interesting observation regarding this (best) configuration is that all the three performance milestones converge for this configuration. That is we obtain the best accuracy and also the best F1 measure (balance between precision and recall) and the default prediction probability cut-off of $0.5$.

\begin{table}[ht]
\scriptsize
\begin{center}
\caption{AE + MLP: Best E2E Configuration (Conf. ID 40)}
\begin{tabular}{|c|c|c|c|}
\hline
\multicolumn{4}{|c|}{\textbf{\textit{Conf. ID 40}}}                   \\ \hline
\multicolumn{1}{|c|}{\textbf{Parameter}} & \multicolumn{1}{c|}{\textbf{Value}} & 
\multicolumn{1}{|c|}{\textbf{Parameter}} & \multicolumn{1}{c|}{\textbf{Value}} \\ \hline
AE Hidden Lyrs. & {[}128,64{]}        &     
MLP Hidden Lyrs.   & {[}64,64,64,64{]}
\\
AE Embedding Lyr.  & 32   & 
MLP Optimizer                   & rmsprop  
\\
AE Batch Size                   & 1024    &          
MLP Batch Size                  & 1024
\\
AE Epochs                       & 1000     &
MLP Epochs                      & 1000 
\\
\hline
AE Val. Loss                    & 0.006    &
MLP AUC                         & 0.814 
\\ 
\hline
Accuracy (Th=0.5)               & 0.772         &
FPR (Th=0.5)                    & 0.11              \\
\hline
BestAccuracy                   & 0.772         &
FPR @ BestAccuracy             & 0.11              \\
\hline
Accuracy @ BestF1              & 0.772             &
FPR @ BestF1                   & 0.11              \\ 
\hline
\end{tabular}
\label{table:best-configuration}
\end{center}
\end{table}

\section{Conclusion} \label{sec:conclusion}
 Even amongst all static features available for Android malware analysis, Implicit Intents are the easiest to extract. In this paper we demonstrate that besides being easy to extract, these Intent are also quite intuitive to analyze descriptively. The results in section \ref{sec:results} also proves that a potent Android IDS could also be created using Implicit Intent alone. In section \ref{sec:experiments} we experimented with over $40$ E2E DL configurations and established a performance benchmark for a malware classifier that uses Implicit Intent as the only features. The benchmark performance had an AUC of $0.814$, along with and accuracy of $77.2\%$ and an FPR of $0.11$. 

\bibliographystyle{IEEEtran}
\bibliography{main}

\end{document}